\documentclass{PoS}

\usepackage{enumitem}

\makeatletter
\newcommand*{\rom}[1]{\expandafter\@slowromancap\romannumeral #1@}
\makeatother

\title{Belle \rom{2} early physics program of bottomonium spectroscopy}

\ShortTitle{Belle \rom{2} early physics program of bottomonium spectroscopy}

\author{\speaker{Hua Ye}\\
        (On behalf of the Belle \rom{2} Collaboration)\\
        DESY, Notkestrasse 85, D-22607 Hamburg, Germany\\
        E-mail: \email{hua.ye@desy.de}}

\abstract{
The Belle \rom{2} experiment at the SuperKEKB collider is a major upgrade of the KEK "$B$ factory" facility in Tsukuba, Japan. Phase 1 commissioning of the main ring of SuperKEKB has started in February 2016 and first physics data will be recorded in 2017 during the so-called Phase~2 commissioning, when the partial Belle \rom{2} detector will be operated still without its vertex detector. In 2018, the full Belle \rom{2} detector will be rolled in and physics run will start. In this proceeding, a possible physics program for this early data run at different center-of-mass energies is described, in particular at the $\Upsilon(3S)$ and $\Upsilon(6S)$ resonances, amongst other energy points.
}

\FullConference{XXIV International Workshop on Deep-Inelastic Scattering and Related Subjects\\
		11-15 April, 2016\\
		DESY Hamburg, Germany}

\begin{document}

\section{Introduction}

The so-called B factory is an asymmetric $e^+e^-$ collider mainly running at the $\Upsilon(4S)$ resonance energy of 10.58 GeV to produce B meson pairs. The first generation of B factories are Belle at KEKB in Japan and BaBar at PEP-\rom{2} in US, which have collected about 1.5 ab$^{-1}$ of data in total (Table.\ref{datasets}). B factories have a fruitful physics program and have made great achievements at the intensity frontier. The physics topics include CP violation in beauty and charm sectors, precise measurements of CKM matrix elements, bottomonium spectroscopy and searching for unanticipated new particles such as XYZ hadrons.
B factories are also competitive in the search for new physics beyond the Standard Model (SM), such as study of the rare decays  $b\to sl^+l^-$, $B\to \tau\nu$ and $B\to D^{(*)}\tau\nu$, lepton flavor violating and searching for light dark matter and dark photons.


The Belle \rom{2} experiment, as the upgraded successor of Belle, is under construction at the SuperKEKB collider~\cite{superkekb}. Benefiting from the nano-beam technology, the designed peak luminosity of SuperKEKB is 8$\times10^{35}$cm$^{-2}$s$^{-1}$. The detector will be upgraded for Belle \rom{2} apparatus, including the vertex detector (VXD) of two-layers DEPFET pixel (PXD) and 4-layers double-sided silicon strips (SVD), a drift chamber (CDC) with longer arms and smaller cells, a completely new particle identification (PID) system, the upgraded electro-magnetic calorimeter (ECL) and $K_L-\mu$ detection system (KLM). As scheduled, the Belle \rom{2} detector will be partially commissioned in the so-called BEAST \rom{2} ({\bf B}eam {\bf E}xorcism for {\bf A ST}able experiment) Phase 2 in 2017. After full commissioning in 2018, Belle \rom{2} is expected to accumulate the equivalent data sets of Belle in one year, and 50 ab$^{-1}$ integrated luminosity by 2025. Data collected at different center-of-mass energies during Phase 2 and the beginning of physics run could yield unique new physics results two years beforehand. Therefore, this period represents an opportunity for the Belle \rom{2} experiment to have an early scientific impact.

In this proceeding, some proposed studies of bottomonium spectroscopy aiming for the early physics program of the Belle \rom{2} experiment are described.

\begin{table}[htbp]
\centering
\caption{Existing $e^+e^-$ datasets collected near $\Upsilon$ resonances.}
\begin{tabular}{|c|c|c|c|c|c|c|} \hline
Experiment & Scans & $\Upsilon(5S)$ & $\Upsilon(4S)$ & $\Upsilon(3S)$ & $\Upsilon(2S)$ & $\Upsilon(1S)$ \\*
& /Off.Res. & 10876MeV & 10580MeV & 10355MeV & 10023MeV & 9460MeV \\*
& fb$^{-1}$ & fb$^{-1}$ 10$^6$ & fb$^{-1}$ 10$^6$  & fb$^{-1}$ 10$^6$  & fb$^{-1}$ 10$^6$  & fb$^{-1}$ 10$^6$ \\*
\hline
CLEO & 17.1 & 0.4 \ 0.1 & 16 \ 17.1 & 1.2 \ 5 & 1.2 \ 10 & 1.2 \ 21 \\*
\hline
BaBar & 54 & $R_b$ scan & 433 \ 471 & 30 \ 122 & 14 \ 99 & - \\*
\hline
Belle & 100 & 121 \ 36 & 711 \ 772 & 3 \ 12 & 25 \ 158 & 6 \ 102 \\*
\hline
\end{tabular}
\label{datasets}
\end{table}

\section{Accelerator/Detector Conditions and Early Physics}

The proposed commissioning of the Belle \rom{2} detector will take place in BEAST \rom{2}, aiming at characterizing the beam-induced backgrounds near the interaction point (IP).
The commissioning scenario will be performed in three stages, they are
\begin{itemize}
\setlength\itemsep{0em}
\item[] Phase 1 (2016.2-6): Beam commissioning without collisions or the Belle \rom{2} detector.
\item[] Phase 2 (2017.11-2018.3): 
Accelerator tuning and evaluation of beam related background. Partial Belle \rom{2} detector will be rolled in without full vertex detector. Collisions will start.
\item[] Phase 3 (2018.10-): Physics run with full Belle \rom{2} detector.
\end{itemize}
During Phase 2, all the outer sub-detectors and one octant of VXD will be present, while the remaining area of VXD will be populated with the non-Belle \rom{2} detectors FANGS, CLAWS and PLUME. Nominal operating energy is at $\Upsilon(4S)$, while a larger energy range from $\Upsilon(1S)$ to 11.25 GeV should be capable. The instantaneous luminosity is expected to reach $10^{34}$cm$^{-2}$s$^{-1}$ and the beam energy spread is expected to be close to the nominal value of about 5 MeV. The majority of the time in Phase 2 will be spent for accelerator commissioning and tuning. If the machine commissioning is accomplished in a timely manner, there will be opportunity for 2 months physics data collection. The estimated integrated luminosity during Phase 2 is about 20 fb$^{-1}$.

During the early running phase, the luminosity will be relatively low, and therefore the triggers could be configured to be looser than at nominal.
The lack of the VXD detector is expected to result in tracking efficiency losses in the low transverse momentum ($p_t$) region since these tracks can not reach or produce sufficient hits to be reconstructed in CDC. Preliminary studies of the photon efficiency indicate that no appreciable difference is expected between Phase 2 and Phase 3 due to nearly equivalent amount of material contributed in VXD area. As a result, early physics analyses relying on photon detection will be as effective in Phase 2 as in Phase 3.

\section{Bottomonium Spectroscopy}

Heavy quarkonium presents an ideal laboratory for testing the interplay between perturbative and nonperturbative QCD~\cite{heavyQuarknium}.
Bottomonium spectroscopy focuses on the existence, quantum numbers, masses and widths of bottomonia states. Of late, progress has occurred mostly at $e^+e^-$ colliders with the capability to obtain large data sets at bottomonia masses with well-known initial-state quantum numbers and kinematics. Although bottomonium spectroscopy has been explored in detail by the first generation of B-factories and other experiments, there are still open questions, for example the unobserved states such as $\Upsilon(1D)$ and some unpredicted observations such as XYZ hadrons. Quite a few of these analyses are limited by statistics. From the view of event reconstruction in the detector, transitions involving another bottomonium state especially the radiative transitions from onresonance data sets do not strongly rely on the vertex determination and PID. All these properties offer achievable opportunities of early physics at Belle \rom{2} in bottomonium spectroscopy.

\section{Bottomonium bellow $\Upsilon(4S)$}

A significant increase in the scientific potential at $\Upsilon(3S)$ could be achieved with about 200 fb$^{-1}$ of data (7$\times$ Babar data size). Within a shorter time of data collection, it offers unique early physics that would not necessarily be achieved with an equivalent size of Belle data at $\Upsilon(4S)$ (711 fb$^{-1}$). Meanwhile, from the standpoint of machine operation, it may be desirable to begin at lower energy.

\subsection{Study of $\eta_b(1S,2S)$ at $\Upsilon(3S)$}

With the 109M $\Upsilon(3S)$ radiative decays, the Babar Collaboration observed the bottomonium ground state $\eta_b(1S)$~\cite{babar_etab}.
Subsequent analysis at Belle from $\Upsilon(5S)$ provided further measurement of $\eta_b(1S)$ and evidence of $\eta_b(2S)$ via $\Upsilon(5S)\to h_b(nP)\pi^+\pi^-\to\eta_b(mS)\gamma$~\cite{belle_etab}. Despite many measurements in different experiments, there is a conflict in the $\eta_b(1S)$ mass at about 3.5$\sigma$ between the combined Babar~\cite{babar_etab,babar_etab2} and CLEO~\cite{cleo_etab} results of 9391.1$\pm$2.9 MeV/$c^2$ from radiative decays and Belle results of 9403.4$\pm$1.9 MeV/$c^2$ from $h_b(nP)\to\gamma\eta_b(1S)$. Further measurements with increased statistics are needed.

Verification of the $\eta_b(1S)$ mass from $\Upsilon(3S)\to\gamma\eta_b(1S)$ 
should be straightforward with a large statistics from early Belle \rom{2} data. Given the branching fraction of ~5$\times10^{-4}$, one expects roughly 800 $\eta_b(1S)$ per fb$^{-1}$, assuming an efficiency of about 40\%. The decay of $\Upsilon(3S)\to\pi^0 h_b(1P)\to\gamma\eta_b(1S)$ for which ~3$\sigma$ evidence was seen at Babar~\cite{babar_etab3} may also provide information on $\eta_b(1S)$. Based on the branching fraction and efficiency from Babar of $4\times10^{-4}$ and ~20\%, this process is expected to represent about 350 events per fb$^{-1}$ at Belle \rom{2}. In both processes $\eta_b(1S)$ are inclusively reconstructed and a large background is expected. Another potential pathway well-suited to the initial running conditions of Belle \rom{2} would be via $\Upsilon(3S)\to\gamma\chi_{b0}(2P)\to\eta\eta_b(1S)$. The branching fraction of $\mathcal{B}(\chi_{b0}(2P)\to\eta\eta_b(1S))$ is expected to be as large as $10^{-3}$~\cite{voloshin}. One can expect about 2000 events with a 200 fb$^{-1}$ $\Upsilon(3S)$ sample assuming the efficiency of 5\%.

\subsection{$\Upsilon(nD)$ studies}

The $\Upsilon(1D)$ is the lowest-lying D-wave triplet of the $b\bar{b}$ system. The CLEO experiment made the first observation of $\Upsilon(1^3D_2)$ using the four-photon cascade of $\Upsilon(3S)\to\gamma\chi_{bJ}(2P)\to\gamma\Upsilon(1D)\to\gamma\chi_{bJ}(1P)\to\gamma\Upsilon(1S)\to l^+l^-$, where $l^{\pm}=e^{\pm}/\mu^{\pm}$~\cite{cleo_Y1D}, and Babar observed $\Upsilon(3S)\to\gamma\gamma\Upsilon(1^3D_2)\to\pi^+\pi^-\Upsilon(1S)$ with the significance of 5.8 $\sigma$~\cite{babar_1d}. Theoretical calculations predict the J=1, 3 masses around 10150, 10170 GeV/$c^2$, the $\Upsilon(2^3D_2)$ mass is predicted to be in the range of 10420 to 10460 MeV/$c^2$. They are expected to be narrow, and predominantly decay to $\gamma\chi_{bJ}(nP)$.
Opportunities are open to Belle \rom{2} with larger statistics $\Upsilon(3S)$ samples. Meanwhile, $\Upsilon(n^3D_1)$ states can be produced directly via a beam energy scan.
Assuming a value of instantaneous luminosity of 2$\times10^{-34}$cm$^{-2}$s$^{-1}$, a 7-10 steps scan centered on 10.150 (10.435) GeV with 2 (1.4) fb$^{-1}$ per point would take about one week for each J=1 states, to achieve a 5$\sigma$ observation.

\subsection{$h_b(1P)$ studies}

First evidence (3.1$\sigma$) of $h_b(1P)$ came from Babar by selecting the soft pion and radiative photon restricted to $\eta_b(1S)$ mass in $\Upsilon(3S)\to\pi^0h_b(1P)\to\gamma\eta_b(1S)$, with the multiplied branching fraction of (4.3$\pm$1.4)$\times10^{-4}$~\cite{babar_etab3}. No evidence of the dipion transition $\Upsilon(3S)\to\pi^+\pi^-h_b(1P)$ was found, the upper limit of the branching fraction was determined to be 1.2$\times10^{-4}$~\cite{babar_hb}. An increase  by a factor of >3 in statistics could provide an observation of $h_b(1P)$. This analysis mainly relies on photon detection, which would be advantageous in early running scenarios of Belle \rom{2}.

\subsection{Analysis with converted photons}

An improvement in photon energy resolution can be achieved using the $e^+e^-$ pair from photon conversion in detector material, although the efficiency for reconstruction is much lower than that for calorimeter. Babar attempted to perform a study of bottomonium radiative transitions using converted photons. Precise measurements of $\mathcal{B}(\chi_{b1,2}(1P,2P)\to\gamma\Upsilon(1S))$ and $\mathcal{B}(\chi_{b1,2}(2P)\to\gamma\Upsilon(2S))$ were made, and the searches for $\eta_b(1S,2S)$ states were inconclusive~\cite{babar_convert}.
The advantage of the improved resolution from a converted photon technique offers chances to make a definitive measurement of $\eta_b(1S)$ mass and width in future B-factories with more data.

\subsection{Hadronic/Radiative transitions}

With sufficient statistics, the dipion transitions amongst bottomonia such as $\Upsilon(3S)\to\pi\pi\Upsilon(1S,2S)$, $\chi_b(2P)\to\pi\pi\chi_b(1P)$ can be studied in detail. Other hadronic transitions that have been attempted include $\Upsilon(2S,3S)\to\gamma\Upsilon(1S)$ and $\chi_b(2P)\to\omega\Upsilon(1S)$. Studies involving $\rho$ transitions need to be done.
%
Regarding these radiative transitions between $\Upsilon$ and $\chi_b$, the $\chi_{b0}(2P)\to\gamma\Upsilon(1S)$ could only reach a significance of 2.2$\sigma$ from recent measurement of Babar, the $\Upsilon(3S)\to\gamma\chi_{bJ}(1P)$ observation is difficult to measure due to the overlapping photon transition energies and theoretically difficult to calculate due to the effects of higher-order corrections.

\section{Bottomonium above $\Upsilon(4S)$}

The upper limit of Belle \rom{2}/Super-KEKB facilities is the center-of-mass energy ($\sqrt{s}$) of 11.25 GeV/$c^2$, which would allow for data collection across $\Upsilon(6S)(\Upsilon(11020))$. The proposed physics program for bottomonia above $\Upsilon(4S)$ includes the energy scan and searches for bottomonium-like states.

\subsection{$\sqrt{s}$ scan}

Scans in $e^+e^-$ $\sqrt{s}$ can map out vector resonances via either inclusive hadronic-event counting ($R$ scan) and/or exclusive final states ($e.g.$, $B\bar{B}$, $\pi\pi\Upsilon(nS)$). Up to date, sparing studies of energy range of 10.6 to 11.25 GeV have been done. In 2008, Babar published the results of $e^+e^-\to b\bar{b}$ cross section measurements based on the 3.3 fb$^{-1}$ data from 10.54 to 11.20 GeV and 600 pb$^{-1}$ at the $\Upsilon(6S)$ region; parameters of $\Upsilon(5S)$ and $\Upsilon(6S)$ were measured~\cite{babar_scan}. Subsequently, Belle measured the production cross section for $e^+e^-\to\Upsilon(nS)\pi^+\pi^-$ $(n=1,2,3)$ using the 8.1 fb$^{-1}$ data between 10.83 and 11.02 GeV, followed by a high-luminosity scan for $e^+e^-\to\Upsilon(nS)\pi^+\pi^-$ and $e^+e^-\to b\bar{b}$~\cite{belle_scan}. For the early running period of Belle \rom{2}, one proposal is doing a scan extending to the energy beyond $\Upsilon(6S)$ or even close to $\Lambda_b$ pair threshold, which is close to the maximum energy of facilities.

\subsection{Charged Bottomonium-like states: $Z_b^{\pm}$}

The bottomonium-like $Z_b^{\pm}(10610)$ and $Z_b^{\pm}(10650)$ states are of special interest since their properties do not fit the potential model predictions. The minimal quark substructure of $b\bar{b}u\bar{d}$ would be necessary and therefore be manifestly exotic.
They were first observed by Belle in $\Upsilon(nS)\pi^{\pm}$ and $h_b(mP)\pi^{\pm}$ ($n=1,2,3;m=1,2$) channels produced in $\Upsilon(5S)\to Z_b^{\pm}\pi^{\mp}$, using the 121 fb$^{-1}$ sample~\cite{belle_zb}. The $J^P=1^+$ is favored from angular analyses. Corresponding neutral state of $Z_b^0(10610)$ produced in $\Upsilon(5S)\to\Upsilon(2S,3S)\pi^0\pi^0$ decays at a consistent mass~\cite{belle_zb0} indicates isospin 1 is favored.

Given proximity to the $B\bar{B}^*$ and $B^*\bar{B}^*$ thresholds and finite widths, it's natural to expect the rates of $Z_b(10610)\to B\bar{B}^*$ and $Z_b(10650)\to B^*\bar{B}^*$ are substantial in the molecular picture.
The $\pi^{\pm}$ missing mass spectrum in $\Upsilon(5S)\to B\bar{B}^*\pi$ decays shows a clear excess (8$\sigma$) of events over background which is interpreted as $Z_b^{\pm}(10610)$ signal, while $\Upsilon(5S)\to B^*\bar{B}^*\pi$ shows $Z_b^{\pm}(10650)$ signal with 6.8$\sigma$~\cite{belle_zb_bb}. Belle measurements indicated that $B^{(*)}\bar{B}^*$ decays are dominant and accounts for a branching fraction of about 80\% assuming so far observed $Z_b$ decays are saturated.

Belle also studied the processes of $e^+e^-\to h_b(1P,2P)\pi^+\pi^-$ with 6 fb$^{-1}$ $\Upsilon(6S)$ sample. Evidence of $Z_b^{\pm}(10610)$ and $Z_b^{\pm}(10650)$ were reported~\cite{belle_zb_6s}, while the information of $Z_b$s on $\pi\Upsilon(mS)$ and $B^{(*)}\bar{B}^*$ from $\Upsilon(6S)$ decay are still absent. Larger statistics $\Upsilon(5S,6S)$ samples are necessary for a better understanding of $Z_b$ states.

\section{Summary}

The Belle \rom{2} experiment will be the next generation of B-factories with ultra high luminosity.
The physics data taking will start in 2018, while first data with a partial detector will come in 2017.
Various bottomonia spectroscopy topics that could be considered in the early phase for Belle \rom{2} are covered. Considering the detector condition, the proposed analyses here are more related to the photon reconstruction but not to PID, nor vertex finding precision. Most of these analyses are limited to existing sample sizes at specific collision energies, especially at $\Upsilon(3S)$ and $\Upsilon(6S)$.



\begin{thebibliography}{99}



\bibitem{superkekb}
T. Abe {\it et al.}, KEK Report 2010-1 (2010), arXiv: 1011.0352v1.
Y. Ohnisi {\it et al.}, PTEP 2013, 03A011.

\bibitem{heavyQuarknium}  N. Brambilla {\it et al.}, Eur. Phys. J. {\bf C71}, 1534 (2011)

\bibitem{babar_etab} B. Aubert {\it et al.}, BaBar Collaboration, Phys. Rev. Lett. {\bf 101} 071801 (2008)

\bibitem{cleo_etab} G. Bonvicini {\it et al.}, CLEO Collaboration, Phys. Rev. D {\bf 81}, 031104 (2010)

\bibitem{belle_etab} R. Mizuk {\it et al.}, Belle Collaboration, Phys. Rev. Lett. {\bf 109}, 232002 (2012)

\bibitem{babar_etab2} B. Aubert {\it et al.}, BaBar Collaboration, Phys. Rev. Lett. {\bf 103} 161801 (2009)

\bibitem{babar_etab3} J.~P.~Less {\it et al.}, BaBar Collaboration, Phys. Rev. D {\bf 84}, 091101 (2011)


\bibitem{voloshin} M. Voloshin, Mod. Phys. Lett. {\bf A19} 2895 (2004)

\bibitem{cleo_Y1D} G. Bonvicini {it et al.}, CLEO Collaboration, Phys. Rev. D {\bf 70} 032001 (2004)

\bibitem{babar_1d} P. del Amo Sanchez {it et al.}, BABAR Collaboration, Phys. Rev. D {\bf 82}, 111102 (2010)

\bibitem{babar_hb} J.~P.~Less {\it et al.}, BaBar Collaboration, Phys.Rev. D{\bf 84} 011104 (2011)

\bibitem{babar_convert} J.~P.~Less {\it et al.}, BaBar Collaboration, Phys.Rev. D {\bf 84} 072002  (2011)

\bibitem{babar_scan} B.~Aubert {\it et al.}, BaBar Collaboration, Phys. Rev. Lett. {\bf 102} 012001 (2009)

\bibitem{belle_scan} K.-F.~Chen {\it et al.}, Belle Collaboration, Phys. Rev. D {\bf 82}, 091106 (2010); D.~Santel {\it et al.}, Belle Collaboration, Phys. Rev. D {\bf 93}, 011101 (2016)

\bibitem{belle_zb} A.~Bodar {\it et al.}, Belle Collaboration, Phys. Rev. Lett. {\bf 108}, 122001 (2012)

\bibitem{belle_zb0} P.~Krokovny {\it et al.}, Belle Collaboration, Phys. Rev. D {\bf 88}, 052016 (2013)

\bibitem{belle_zb_bb} I.~Adachi {\it et al.}, Belle Collaboration, arXiv:1209.6450 [hep-ex]

\bibitem{belle_zb_6s} I.~Adachi {\it et al.}, Belle Collaboration, arXiv:1508.06562 [hep-ex]

\end{thebibliography}
\end{document}